\documentclass[journal]{IEEEtran}

\usepackage{graphicx, subfigure}
\usepackage{amsmath}
\usepackage[colorlinks,linkcolor=blue]{hyperref}
\usepackage{cite}
\usepackage{amsfonts} 
\usepackage{multirow}
\usepackage{booktabs}  

\ifCLASSINFOpdf

\else

\fi

\begin{document}

\title{STCC---A Unified Source-Channel Semantic Token Coding Framework for Semantic Communications}

\author{Zhicheng~Bao,~Chen~Dong$^*$,~Sen~Wang,~Long~Liu,~Nan~Ma,~\IEEEmembership{Member,~IEEE},~Hao~Chen,~\IEEEmembership{Member,~IEEE},~Xiaodong~Xu,~\IEEEmembership{Senior~Member,~IEEE},~Yinqiu Liu,~Ping~Zhang,~\IEEEmembership{Fellow,~IEEE}

\thanks{This work is supported in part by the Beijing Natural Science Foundation under Grant L251035, in part by the National Natural Science Foundation of China No. 62401303, in part by the National Science and Technology Major Project-Mobile Information Networks under Grant No. 2024ZD1300700, in part by Beijing University of Posts and Telecommunications-China Mobile Communications Group Co., Ltd. Joint Institute. (\textit{Corresponding authors: Chen Dong})}

\thanks{Zhicheng Bao, Nan Ma and Xiaodong Xu are with the State Key Laboratory of Networking and Switching Technology, Beijing University of Posts and Telecommunications, Beijing 100876, China, and also with the Department of Broadband Communication, Peng Cheng Laboratory, Shenzhen 518000, China. (e-mail: zhicheng\_bao@bupt.edu.cn; manan@bupt.edu.cn; xuxiaodong@bupt.edu.cn)}

\thanks{Chen Dong, and Ping Zhang are with the State Key Laboratory of Networking and Switching Technology, Beijing University of Posts and Telecommunications, Beijing 100876, China. (e-mail: dongchen@bupt.edu.cn; pzhang@bupt.edu.cn).}

\thanks{Long Liu and Hao Chen are with the Department of Broadband Communication, Peng Cheng Laboratory, Shenzhen 518000, China (email: liul05@pcl.ac.cn; chenh03@pcl.ac.cn).}

\thanks{Sen Wang is with China Mobile Research Institute, Beijing, China. (email: wangsenyjy@chinamobile.com)}

\thanks{Yinqiu Liu is with the College of Computing and Data Science, Nanyang Technological University, Singapore
(email: yinqiu001@e.ntu.edu.sg).}

}

\markboth{Journal of \LaTeX\ Class Files,~Vol.~XX, No.~XX, XXX~2023}%
{Shell \MakeLowercase{\textit{et al.}}: Bare Demo of IEEEtran.cls for IEEE Journals}

\maketitle
\begin{abstract}
Deep Joint Source-Channel Coding (JSCC) has emerged as a promising paradigm for overcoming the ``cliff effect" in wireless communications. However, existing Deep JSCC frameworks operate directly on raw analog data such as image pixels rather than the discrete semantic tokens that foundation models require. Moreover, traditional systems employ fixed, hand-designed constellations that treat all tokens equally, leading to catastrophic random errors under channel noise. In this paper, the Semantic Token Codebook Communication (STCC) is proposed as a unified source-channel semantic token coding framework designed to transmit the discrete semantic tokens of foundation models over noisy channels. The core of STCC is the Semantic Token Codec (STC). It accepts discrete tokens as input, which maintains compatibility with foundation models while employing a residual multiple layer perceptron, i.e., MLP-based encoder that learns geometrically structured constellations optimized with a triple-loss objective. This learned mapping forces the channel topology to align with the semantic embedding space, ensuring that channel noise results in topological errors rather than random corruption. This phenomenon is theoretically and empirically characterized, identifying ``Semantic Drift" in symbolic modalities and ``Structural Distortion" in perceptual modalities, where errors shift predictions to semantically or structurally similar tokens. Extensive experiments demonstrate that STCC significantly outperforms traditional systems in low-SNR regimes, effectively converting channel noise into semantic variations without requiring receiver-side modification.

\end{abstract}

\begin{IEEEkeywords}
    Semantic communication, semantic token codebook communication, semantic token coding, semantic drift, structural distortion, generalized framework
\end{IEEEkeywords}

\IEEEpeerreviewmaketitle
\section{Introduction}
\IEEEPARstart{S}{emantic} communications are evolving as a beyond-Shannon-type communication paradigm, expected to be a pivotal technology for future wireless systems \cite{9679803, Niu2022APS}. Unlike traditional systems that focus on the accurate transmission of bit sequences, semantic communications prioritize the conveying of meaning or the effectiveness of task execution.

Recently, the rapid proliferation of foundation models, such as Large Language Models (LLMs) and Large Vision Models (LVMs), has fundamentally transformed this landscape. These systems have demonstrated unprecedented capabilities in understanding and generating content across diverse modalities. A critical, unifying characteristic of these modern generative AI systems is their reliance on a standardized data interface: the discrete semantic token. Whether representing a sub-word unit in text via tokenizers like WordPiece in BERT \cite{Devlin2019BERTPO}, a spatial patch in an image via Vector Quantized networks like VQ-GAN \cite{Esser2020TamingTF} or MaskGIT \cite{Chang2022MaskGITMG}, tokens are the atomic, discrete units that transformers process to capture underlying semantic patterns.

However, a critical disconnect exists between the native discrete format of foundation models and current wireless infrastructure. Traditional systems, relying on the separation principle, treat token indices as arbitrary bit sequences. While effective for data integrity, they suffer from a sharp ``cliff effect,'' where slight channel noise causes random bit errors that flip a token to a completely unrelated index (e.g., using LDPC \cite{Richardson2018DesignOL}). 

To address this, Deep Joint Source-Channel Coding (JSCC) has emerged as a paradigm shift, utilizing neural networks to map data directly to continuous symbols for graceful degradation \cite{Fresia2010JointSA, Guyader2001JointST}. Yet, a critical gap remains: existing JSCC methods predominantly operate on raw analog data (e.g., pixels) rather than discrete tokens, creating an integration barrier with pre-trained foundation models. Meanwhile, conventional systems that do accept discrete tokens employ fixed constellations (e.g., QPSK) that lack semantic awareness, inevitably leading to random semantic corruption.

There is a pressing need for a unified communication framework that is natively compatible with discrete tokens while being semantically aware to inherit the robustness benefits of JSCC. Such a system should accept discrete tokens from foundation models, map them to learned channel constellations that preserve semantic structure, and ensure that inevitable channel corruption results in soft, meaningful variations rather than catastrophic random failures.

\section{Related Works}
\subsection{Deep Joint Source-Channel Coding}
Deep JSCC revolutionizes wireless transmission by learning direct source-to-channel mappings, effectively bypassing the separation principle \cite{Bourtsoulatze2019DeepJS, Xie2020DeepLE}. This paradigm has been successfully adapted across diverse modalities, including text \cite{8461983}, speech \cite{10477917}, images \cite{Yang2021DeepJS, Dong2023SemanticCS}, video \cite{10685066}, and 3D point clouds \cite{10679082}, demonstrating superior robustness against the ``cliff effect.'' Beyond data reconstruction, task-oriented systems have been developed to minimize latency by transmitting only features relevant for specific tasks like transcription \cite{10889160} or control \cite{10445833}, with multi-task architectures \cite{Wang2021DeepJS, Jankowski2020WirelessIR, Lo2023CollaborativeSC} further optimizing efficiency. However, these methods predominantly operate on raw continuous data, preventing direct interfacing with the discrete semantic tokens of modern foundation models.

\subsection{Discrete Semantic Communications}
To bridge the gap between semantic robustness and digital infrastructure, recent research addresses the mismatch between analog features and discrete modulation. Several approaches quantize latent representations to predefined constellations \cite{9838671, 10304507, 10495330}, while others employ non-linear quantization \cite{10200355} or semantic-aware frameworks like sDAC \cite{10985906}. Advanced schemes integrate OFDM \cite{10521803}, mixed analog-digital architectures \cite{Dai2021CommunicationBT}, or channel-adaptive strategies \cite{10584091}. While successful in digitizing transmission, these methods typically rely on fixed constellations (e.g., QPSK) that ignore semantic geometry. In contrast, our proposed STC explicitly learns the topological structure of the constellation, ensuring that physical channel geometry aligns with the semantic embedding space.

\subsection{Semantic Communications for Foundation Models}
The advent of foundation models has spurred interest in Token Communications, a paradigm formally introduced by \cite{11175596}, where the discrete token serves as the fundamental unit of information exchange. This framework leverages the cross-modal context of pre-trained MLLMs to enable semantic source compression, channel coding, and network protocols that reduce retransmissions by predicting lost tokens.

Building on this concept, recent research has focused on optimizing the efficiency of token transmission. Devoto \textit{et al.} \cite{Devoto2025AdaptiveST} proposed an adaptive token communication scheme for edge inference, introducing a mechanism to dynamically identify and transmit only task-relevant tokens based on bandwidth constraints, discarding less informative ones. In the domain of multiple access, Qiao \textit{et al.} \cite{Qiao2025ToDMALM} introduced Token-Domain Multiple Access (ToDMA). By exploiting the ``semantic orthogonality'' of tokens from different users, ToDMA allows simultaneous transmission over shared resources, utilizing receiver-side LLMs to separate collided signals and achieve latency gains over orthogonal schemes.

Parallel efforts have addressed error resilience through packetization and auxiliary guidance. In \cite{11143259}, the SemPA-GBeam algorithm was developed, which uses genetic beam search to optimize token packetization, ensuring that packet losses over erasure channels result in minimal semantic similarity degradation. Taking a different approach, \cite{11149073} proposed a text-guided image transmission scheme. By transmitting text descriptions as robust auxiliary information and utilizing standard 5G NR Polar codes with CRC, their receiver detects corrupted tokens and repairs them using a text-conditioned generative transformer MaskGen.

\subsection{Motivations and Contributions}
To address the limitations of existing works, which typically rely on fixed constellations (e.g., Polar codes \cite{11149073}) or expensive receiver-side generative repair, we propose the STCC. STCC is a unified framework designed to bridge the gap between discrete foundation model tokens and continuous channel transmission. The core STC employs a residual MLP to learn geometrically structured constellations where physical proximity reflects semantic similarity. Unlike generic mappings, STC aligns the channel topology with the semantic embedding space, ensuring that channel noise results in meaningful topological errors, specifically ``Semantic Drift'' in symbolic modalities and ``Structural Distortion'' in perceptual modalities, rather than catastrophic failure.

The primary contributions of this paper are summarized as follows:

(1) \textbf{Semantic Token Codebook Communication (STCC) Framework}: A generalized system model is introduced that treats the semantic token as the universal interface, unifying the transmission of different modalities into a single discrete coding paradigm. This approach achieves Deep JSCC robustness while maintaining full compatibility with foundation model token representations.

(2) \textbf{Semantic Token Codec (STC)}: A novel and highly efficient discrete source-channel coding framework is proposed to transmit foundation model tokens over noisy channels. It employs a residual MLP-based mapper optimized with a specific triple loss objective, which learns a geometrically structured continuous constellation where the physical channel symbols inherit the high-dimensional semantic topology of the source embeddings.

(3) \textbf{Topological Error Characterization}: Two novel phenomena are theoretically and empirically characterized: Semantic Drift in symbolic modalities (e.g., text), where errors shift to synonymous tokens, and Structural Distortion in perceptual modalities (e.g., images), where errors shift to perceptually similar patches. This analysis proves that STC successfully converts random channel noise into topologically bounded semantic variations.

(4) \textbf{Performance Validation}: The efficacy of the proposed STC is validated on standard benchmarks. Experimental results demonstrate that STC significantly outperforms traditional separated communication systems and existing semantic Deep JSCC baselines in low-SNR regimes, validating its ability to maintain high semantic fidelity through learned topological error shaping.

 \begin{figure*}
  \centerline{\includegraphics[width=1\textwidth]{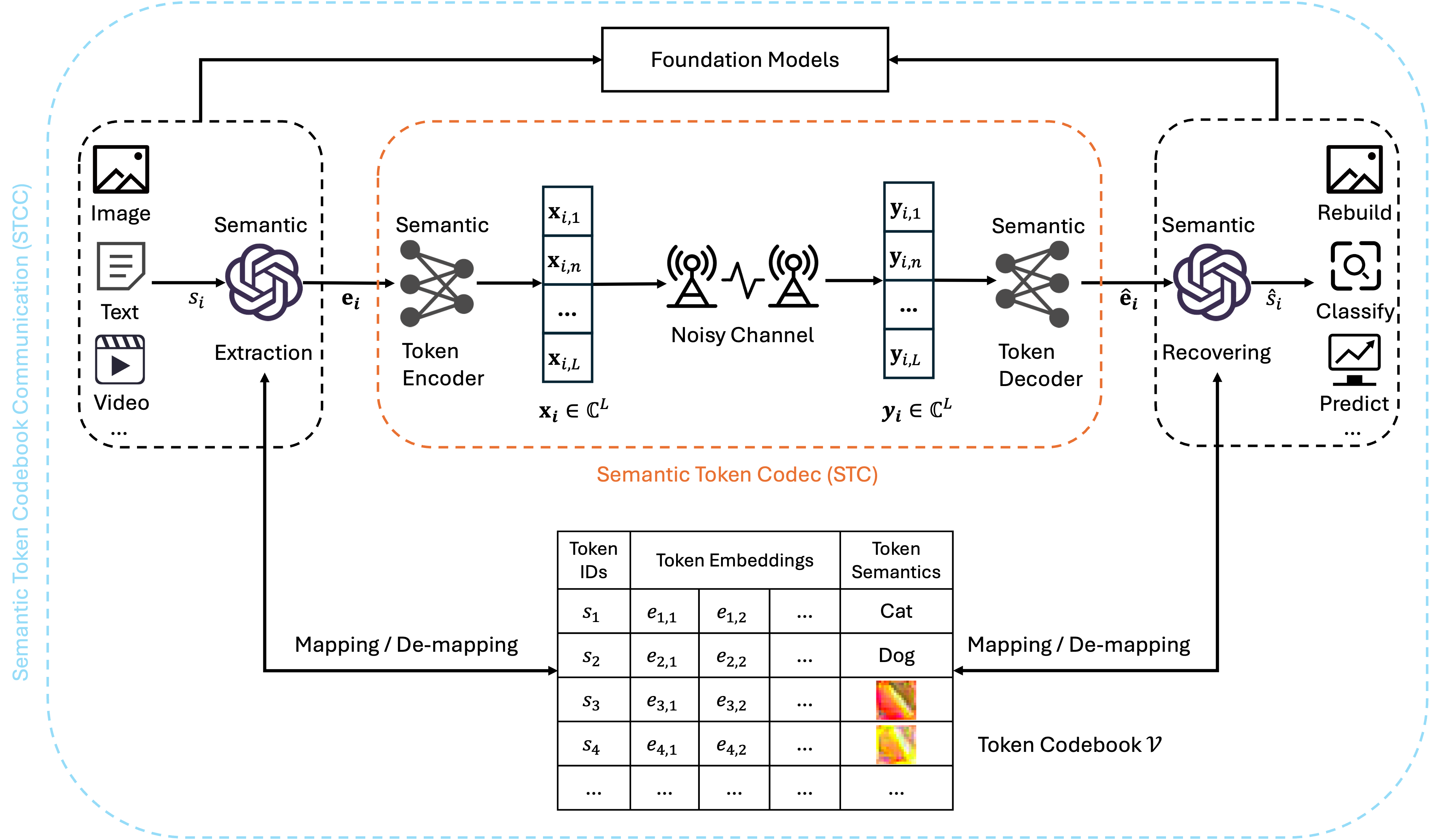}}
  \caption{System model of the proposed STCC. At both sides are the semantic extraction and recovery framework driven by Foundation Models. Between them are the STC transceiver architecture, which transforms semantic embeddings $\mathbf{e}_i$ into learned continuous channel symbols $\mathbf{x}_i$ for transmission over a noisy channel and recovers them as $\hat{\mathbf{e}}_i$. The shared Token Codebook $\mathcal{V}$ is embedded in the system, serving as a unified interface that maps discrete Token IDs to their continuous Token Embeddings and specific semantic meanings.
  \label{system model}}
\end{figure*}

\section{System Model}
In this section, the unified system model of the proposed STCC and STC is presented. First, an architectural overview of STCC is provided. Then, the physical channel model used for transmission is specified. Following this, the end-to-end communication process from discrete token generation to semantic reconstruction is formalized. Finally, the generalized optimization goal that guides the training of the system is defined.

\subsection{STCC Architecture}
The architecture of the STCC is illustrated in Fig.~\ref{system model}. It is composed of three primary blocks: the Foundation Models, the STC Module, and the Noisy Channel.

The Foundation Models represent any pre-trained token-based model, e.g., BERT for text or VQ-GAN for images, that processes information in the form of discrete semantic tokens. Unlike traditional systems that view these tokens as arbitrary data to be compressed, the system treats them as the fundamental units of semantic information. Crucially, these models interface with the system via a shared Token Codebook, which is shown at the bottom of Fig.~\ref{system model}. It acts as a unified dictionary mapping discrete indices to continuous semantic representations.

The STC Module serves as the core interface between the discrete semantic domain and the continuous physical channel domain. It consists of a Semantic Token Encoder at the transmitter and a Semantic Token Decoder at the receiver. The encoder employs a learned mapping that transforms high-dimensional discrete token embeddings into geometrically structured continuous channel symbols, where the constellation topology reflects the semantic relationships in the embedding space. The decoder recovers the semantic features from noisy received signals and maps them back to discrete tokens. This design creates a unified discrete token interface with learned semantic-aware continuous constellations, making it simultaneously compatible with foundation models and robust to channel noise.

\subsection{Channel Model}
To simulate the physical transmission environment, the wireless link is modeled as a continuous-valued channel. Let $\mathbf{x}_i \in \mathbb{C}^{L}$ denote the complex-valued symbol vector generated by the STC encoder for a given $i$-th semantic token, where $L$ represents the number of channel uses per token, which defines the bandwidth expansion ratio.

The received signal $\mathbf{y} \in \mathbb{C}^{L}$ at the destination is modeled as:
\begin{equation}
    \mathbf{y} = h\mathbf{x} + \mathbf{n},
    \label{eq:channel_model}
\end{equation}
where $h$ represents the channel gain coefficient, and $\mathbf{n} \sim \mathcal{N}(0, \sigma_n^2 \mathbf{I}_L)$ denotes the additive white Gaussian noise vector with variance $\sigma_n^2$.

The transmitted symbols are subject to an average power constraint to maximize energy efficiency:
\begin{equation}
    \frac{1}{L} \mathbb{E}[||\mathbf{x}||_2^2] \le P,
    \label{eq:power_constraint}
\end{equation}
where $P$ is the maximum allowable average power. The channel condition is quantified by the SNR, defined as $\text{SNR} = 10\log_{10}(\frac{P}{\sigma_n^2})$ dB.

\subsection{End-to-End Communication Process}
The entire end-to-end communication process can be mathematically formalized as a sequence of transformations mapping the source token space to the reconstructed token space.

Let the source sequence be $S = [s_1, s_2, \dots, s_K]$, where each $s_i \in \mathcal{V}$ is a discrete index selected from a standardized token codebook $\mathcal{V}$. As detailed in Fig. \ref{system model}, this codebook acts as a lookup table containing aligned triples: the discrete Token ID $s_i$, the continuous Token Embeddings, and the corresponding Token Semantics.

First, the foundation model's embedding layer $E(\cdot)$ accesses this codebook to map the discrete index $s_i$ into a dense semantic vector $\mathbf{e}_i \in \mathbb{R}^D$:
\begin{equation}
    \mathbf{e}_i = E(s_i),
\end{equation}
where $D$ denotes the embedding dimension.

The semantic token encoder, denoted as $f_\theta(\cdot)$, learns a mapping from this embedding to a continuous channel symbol vector $\mathbf{x}_i$. Unlike fixed constellations that assign symbols arbitrarily, $f_\theta$ is optimized to preserve the semantic topology of the embedding space:
\begin{equation}
    \mathbf{x}_i = f_\theta(\mathbf{e}_i),
\end{equation}
where $\theta$ represents the trainable parameters of the encoder.

After transmission through the noisy channel defined in Eq.~\eqref{eq:channel_model}, the receiver obtains the noisy vector $\mathbf{y}_i$. The semantic token decoder, denoted as $g_\phi(\cdot)$, attempts to recover the original semantic embedding:
\begin{equation}
    \hat{\mathbf{e}}_i = g_\phi(\mathbf{y}_i),
\end{equation}
where $\phi$ represents the trainable parameters of the decoder.

Finally, the recovered discrete token $\hat{s}_i$ is determined by searching the codebook $\mathcal{V}$ for the embedding that maximizes the cosine similarity with the estimated embedding $\hat{\mathbf{e}}_i$. This mapping operation, denoted as $E^{-1}$, is formalized as:
\begin{equation}
    \hat{s}_i = \underset{k \in \mathcal{V}}{\arg\max} \frac{\hat{\mathbf{e}}_i \cdot E(k)}{\|\hat{\mathbf{e}}_i\|_2 \|E(k)\|_2},
    \label{eq:cosine}
\end{equation}
where $E(k)$ represents the reference embedding for the $k$-th token in the codebook. The end-to-end chain is thus:
\begin{equation}
    s_i \xrightarrow{E} \mathbf{e}_i \xrightarrow{f_\theta} \mathbf{x}_i \xrightarrow{\text{Channel}} \mathbf{y}_i \xrightarrow{g_\phi} \hat{\mathbf{e}}_i \xrightarrow{E^{-1}} \hat{s}_i.
\end{equation}

\subsection{Optimization Goal}
The fundamental goal of the STCC is to maximize the semantic fidelity of the reconstructed tokens under varying channel conditions. Unlike traditional communication systems that minimize Bit Error Rate (BER) with fixed constellations, STCC jointly learns the constellation mapping and semantic recovery to minimize the semantic distance between the transmitted embedding $\mathbf{e}_i$ and the received embedding $\hat{\mathbf{e}}_i$.

Formally, the system is trained to minimize a composite objective function $\mathcal{L}_{STC}$ over the parameters $\theta$ and $\phi$:
\begin{equation}
    (\theta^*, \phi^*) = \arg\min_{\theta, \phi} \mathbb{E}_{s \sim \mathcal{S}, h, \mathbf{n}} [\mathcal{L}_{STC}(\mathbf{e}, \hat{\mathbf{e}})].
    \label{eq:optimization_goal}
\end{equation}
The loss function $\mathcal{L}_{STC}$ is designed to enforce geometric consistency in the channel space, ensuring that the learned constellation aligns with semantic structure such that channel noise results in topological shifts to semantically similar tokens rather than catastrophic random errors. The specific composition of this loss function is detailed in Section IV.

\section{Proposed Method: Topology-Preserving Token Codec}
To bridge the gap between discrete foundation models and continuous physical channels, the STC learns a topology-preserving mapping using two core modules: a residual MLP-based encoder and a Transformer-based decoder. This is a simple yet effective design that balances model capacity and computational efficiency.

\subsection{Semantic Token Encoder}
The encoder $f_\theta(\cdot)$ maps the high-dimensional embedding $\mathbf{e}_i \in \mathbb{R}^D$ to complex channel symbols $\mathbf{x}_i \in \mathbb{C}^L$. To capture non-linear semantic geometry, it employs a stack of $N_b$ residual blocks (Linear - LN - GELU - Linear - LN) following an initial projection to hidden dimension $d_h$. A final linear layer adjusts the output to $2L$, representing real and imaginary components. Crucially, a parameter-free Layer Normalization is applied last to strictly enforce the average power constraint $P$ in Eq.~\eqref{eq:power_constraint} by normalizing the output to unit variance.

\subsection{Semantic Token Decoder}
Since semantic meaning is context-dependent, the decoder $g_\phi(\cdot)$ employs a Transformer \cite{Vaswani2017AttentionIA} to recover tokens from the noisy sequence $\mathbf{Y}$. The input symbols $\mathbb{C}^L$ are concatenated into $\mathbb{R}^{2L}$, projected to model dimension $d_m$, and augmented with positional embeddings. Self-attention layers then exploit global context to denoise the sequence, producing refined embeddings $[\hat{\mathbf{e}}_1, \dots, \hat{\mathbf{e}}_K]$ for the nearest-neighbor retrieval defined in Eq.~\eqref{eq:cosine}.

\subsection{Topological Shaping via Triple Loss Function}
The core innovation of the proposed STC is its ability to learn a channel constellation whose geometry mirrors the semantic topology of the source data. This addresses a fundamental geometric mismatch between the physical and semantic domains:
\begin{enumerate}
    \item The Channel Noise Geometry: Under the AWGN assumption, the additive noise perturbation is isotropic, meaning the error vector $\mathbf{n}$ is equally likely to displace the transmitted symbol in any direction within the high-dimensional hypersphere.
    \item The Semantic Manifold: The source data lies on a complex, non-linear manifold where directions are not uniform. Euclidean proximity does not strictly equate to semantic similarity due to the ``hubness" phenomenon, where certain generic tokens become universal nearest neighbors in the embedding space.
\end{enumerate}

Fixed constellations assign symbols arbitrarily without regard to semantic structure, causing random channel perturbations to produce catastrophic semantic errors. Standard MSE-based autoencoders often fail to preserve the complex topology of semantic embeddings, leading to ``semantic collapse" where small, random channel perturbations push the signal into undefined regions of the semantic space. Conversely, using only Cross-Entropy (CE) ignores the geometric layout entirely, failing to shape the learned constellation structure.

To enforce topology-preserving transmission, a Triple Loss Function is introduced that simultaneously optimizes for geometric fidelity, directional alignment, and classification decisiveness. The total loss $\mathcal{L}_{\text{STC}}$ is defined as:
\begin{equation}
    \mathcal{L}_{\text{STC}} = \mathcal{L}_{\text{CE}} + \lambda_1 \mathcal{L}_{\text{MSE}} + \lambda_2 \mathcal{L}_{\text{Cos}},
\end{equation}
where $\lambda_1$ and $\lambda_2$ are weighting hyperparameters.

\subsubsection{Cross-Entropy Loss ($\mathcal{L}_{\text{CE}}$) -- The Semantic Anchor}
While the channel transmission uses continuous symbols, the recovery target is discrete. $\mathcal{L}_{\text{CE}}$ acts as the semantic anchor, forcing the decoder's output to align with the specific decision boundaries of the correct token index $s_i$. 

It operates on the logits $\mathbf{z}_i = \mathbf{W}_{emb}^T \hat{\mathbf{e}}_i$, where $\mathbf{W}_{emb}$ is the fixed vocabulary embedding matrix:
\begin{equation}
    \mathcal{L}_{\text{CE}} = -\sum_{i=1}^{K} \log \frac{\exp(\mathbf{z}_{i, s_i} / \tau)}{\sum_{j \in \mathcal{V}} \exp(\mathbf{z}_{i, j} / \tau)}.
\end{equation}
Here, $\tau$ is a temperature parameter ($\tau < 1$) used to sharpen the probability distribution. In high-dimensional spaces, a standard Softmax ($\tau=1$) often results in a smeared probability mass across many semantically related neighbors. This allows the decoder to cheat by producing a reconstructed embedding $\hat{\mathbf{e}}_i$ that sits in the safe geometric center of a cluster of tokens. For example, the average of ``cat," ``dog," and ``pet", which can be a meaningless embedding rather than pointing to a specific valid token. By lowering $\tau$, a hard argmax operation is approximated during training, penalizing these ambiguous average representations and forcing the encoder to map the signal towards the precise centroid of the target token $s_i$ in the learned constellation.

\subsubsection{Mean Squared Error Loss ($\mathcal{L}_{\text{MSE}}$) -- Global Regularization}
MSE forces the reconstructed vector $\hat{\mathbf{e}}_i$ to be Euclidean-close to the original embedding $\mathbf{e}_i$. Its primary role is to constrain the global energy scale of the learned constellation:
\begin{equation}
    \mathcal{L}_{\text{MSE}} = \frac{1}{K} \sum_{i=1}^{K} ||\mathbf{e}_i - \hat{\mathbf{e}}_i||_2^2.
\end{equation}
Since the magnitude of a semantic vector often encodes feature intensity, $\mathcal{L}_{\text{MSE}}$ prevents the exploding embedding phenomenon where the encoder artificially inflates signal power to overcome noise, ensuring the system respects the power constraints defined in Eq.~\eqref{eq:power_constraint}.

\subsubsection{Cosine Embedding Loss ($\mathcal{L}_{\text{Cos}}$) -- Directional Alignment}
Modern foundation models (e.g., BERT, CLIP) predominantly rely on the angle between vectors to encode semantic relationships. Two tokens are semantically similar if their vectors share the same direction, regardless of magnitude.

To capture this, $\mathcal{L}_{\text{Cos}}$ explicitly minimizes the angular deviation:
\begin{equation}
    \mathcal{L}_{\text{Cos}} = \frac{1}{K} \sum_{i=1}^{K} (1 - \text{sim}(\mathbf{e}_i, \hat{\mathbf{e}}_i)),
\end{equation}
where $\text{sim}(\mathbf{u}, \mathbf{v}) = \frac{\mathbf{u}^T \mathbf{v}}{||\mathbf{u}|| ||\mathbf{v}||}$. This term is crucial for robustness in the learned constellation; it shapes the noise tolerance regions into cones rather than spheres. This ensures that even if channel noise alters the magnitude of the received vector, which happens frequently in fading channels, the semantic identity encoded in the angle is preserved.

By jointly optimizing these three objectives, STC achieves topological shaping: it learns a channel constellation where the isotropic Gaussian noise balls in the physical channel map onto semantically tolerant regions in the embedding space. Unlike fixed constellations that cause arbitrary errors, the learned mapping ensures that even if the signal is perturbed, the decoded token $\hat{s}_i$ effectively drifts to a synonym rather than an unrelated concept.

\begin{figure*}
  \centerline{\includegraphics[width=0.95\textwidth]{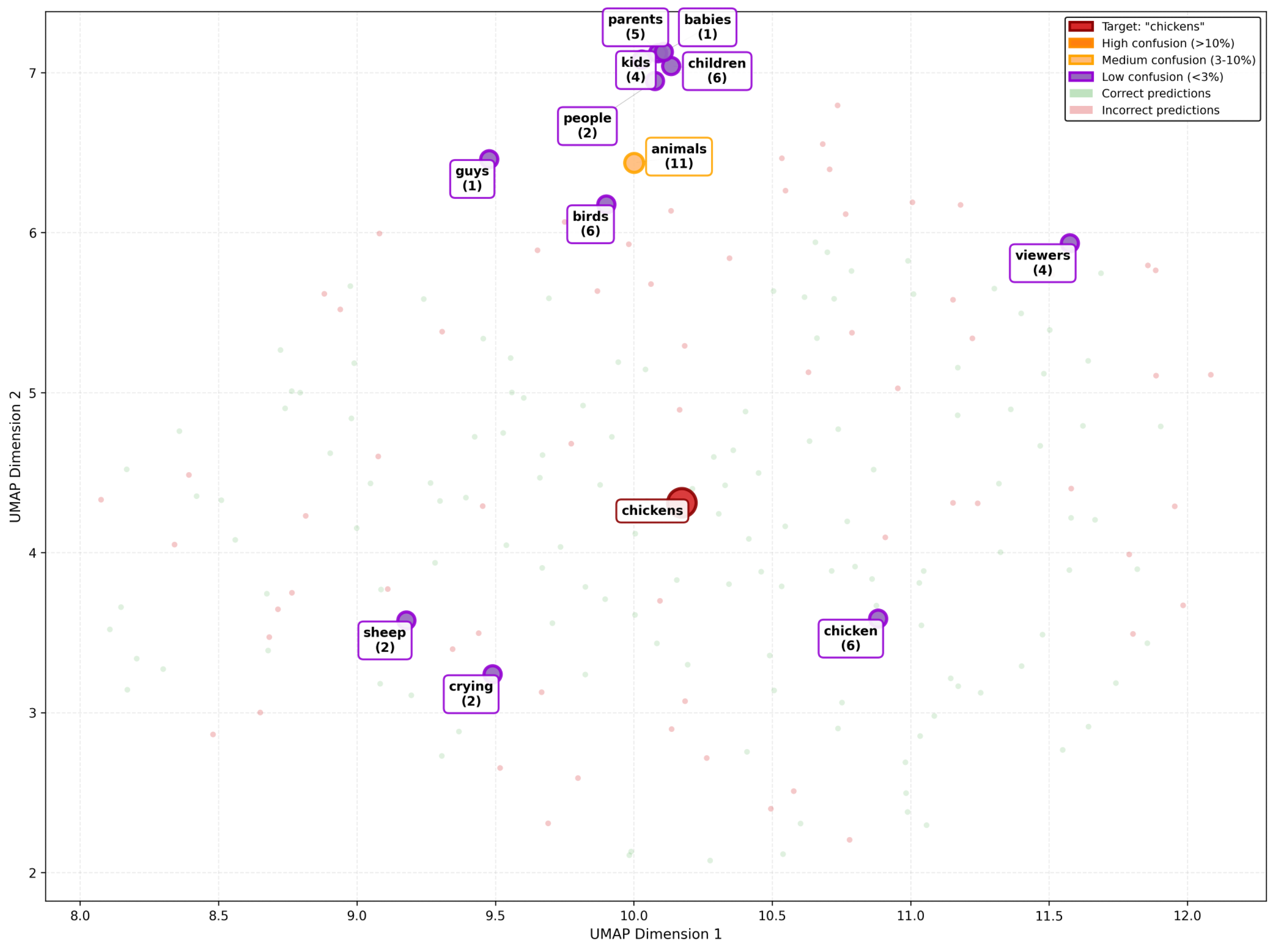}}
  \caption{UMAP projection of learned channel symbols for the target word ``chickens'' and its confusion words under 200 times noisy channel transmission (SNR = 0~dB, $L=32$). Each point represents a confusion word's high-dimension constellation, with labels showing confusion frequency. The color coding indicates confusion rate categories, along with correct predictions (green) and other incorrect predictions (pink). The target word ``chickens'' (red) clusters with semantically related animal terms (``animals'', ``chicken'', ``birds'') in the lower region, while semantically distant words (``children'', ``parents'', ``babies'', ``kids'') form a separate cluster in the upper region, demonstrating that learned constellation topology encodes semantic relationships.}
  \label{text drift}
\end{figure*}

\begin{figure}
  \centerline{\includegraphics[width=0.45\textwidth]{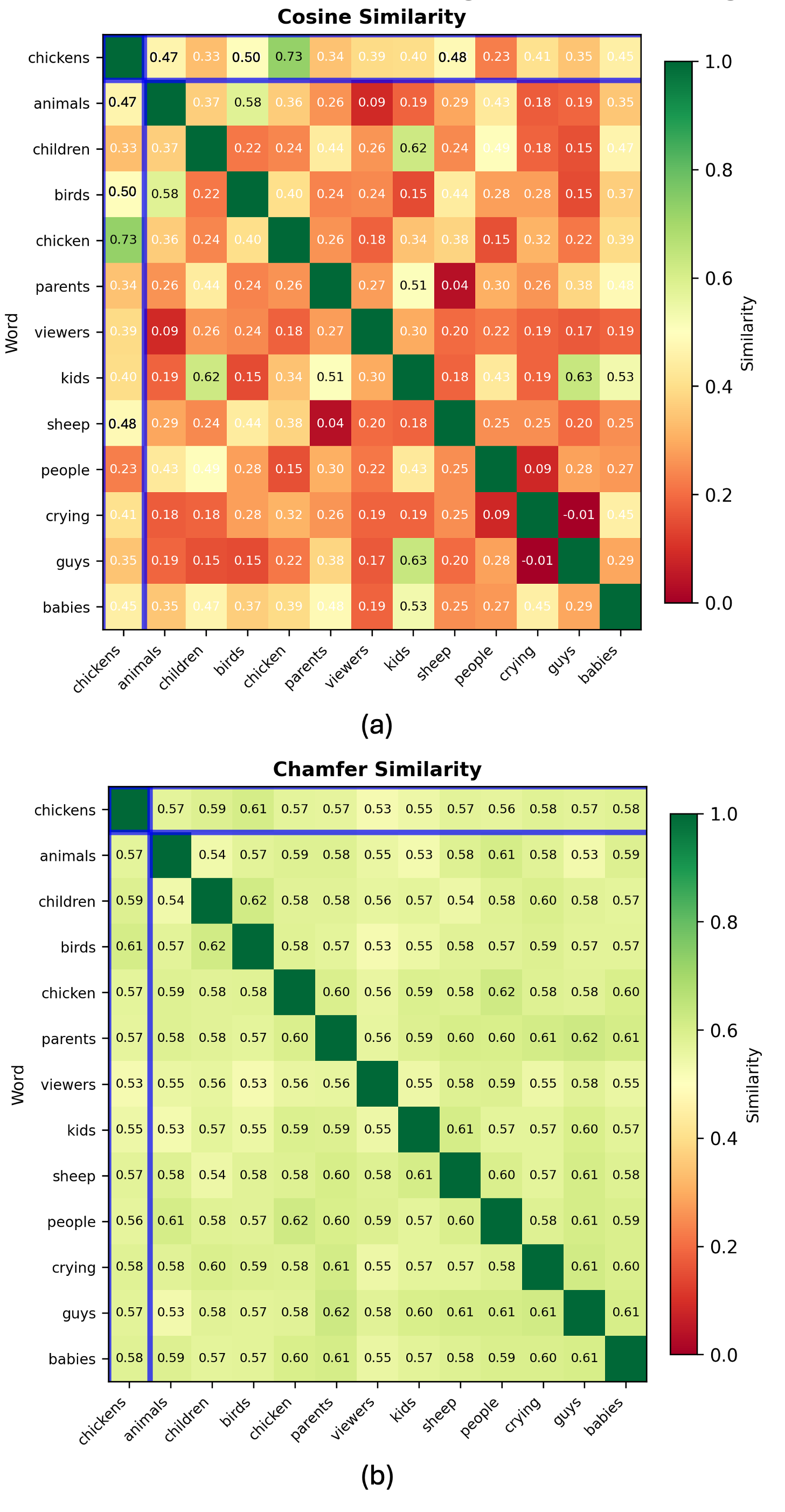}}
  \caption{Similarity matrices between target word ``chickens'' and confusion words showing (a) cosine similarity and (b) chamfer similarity. The target word exhibits high cosine similarity with semantically related words, such as ``animals'': 0.47, ``chicken'': 0.73, ``birds'': 0.50, ``sheep'': 0.48, while showing lower similarity with unrelated words. Chamfer similarity values remain relatively uniform (0.53--0.61), confirming that STC encodes semantic structure through angular relationships rather than simple geometric shape clustering.}
  \label{text metric}
\end{figure}

\section{Topological Analysis of Channel Errors}
We classify modalities into two broad categories based on the nature of their data representation: Symbolic Modalities and Perceptual Modalities. Symbolic modalities, such as text, code, or knowledge graphs, consist of discrete, abstract symbols where meaning is defined by semantic and syntactic relationships rather than physical similarity. Perceptual modalities, such as images, audio, and video, consist of continuous signals (though discretized for processing) where meaning is derived from structural, spatial, or temporal patterns. The proposed STC framework demonstrates distinct but topologically consistent error behaviors for both categories.

\subsection{Theoretical Basis: Angular Topology Determines Error Type}
For analytical tractability, the theoretical analysis focuses on the AWGN channel ($h=1$ in Eq.~\eqref{eq:channel_model}), which isolates the fundamental topological properties of the learned constellation without the additional complexity of fading statistics. The key insights that angular proximity on the learned hypersphere determines error patterns generalize to fading channels, as empirically validated in Section VI. Under fading, the effective noise is scaled by $1/|h|$, but the angular relationships between constellation points remain preserved through the decoder's normalization operations.

In traditional communication systems, semantic tokens are mapped to arbitrary bit sequences through fixed constellations. A single bit flip during transmission (e.g., $1011 \to 1010$) causes a jump to a random vocabulary index. Since standard constellation assignments are semantically arbitrary, the erroneous token is statistically likely to be orthogonal in meaning to the original, e.g., replacing ``chickens'' with ``banana'', resulting in catastrophic semantic failure.

In contrast, STC learns a semantic-aware constellation $\mathcal{C} = \{\mathbf{c}_1, \dots, \mathbf{c}_V\} \subset \mathbb{C}^L$, where each $\mathbf{c}_i = f_\theta(\mathbf{e}_i)$ represents the channel symbol for token $s_i$. Through the power normalization enforced by LayerNorm in the encoder's final layer, all channel symbols are constrained to a hypersphere:
\begin{equation}
    ||\mathbf{c}_i||_2^2 = L \quad \forall i \in \{1, \dots, V\},
\end{equation}
where $L$ is the number of complex channel symbols per token.

On this hypersphere, the relevant distance metric is angular rather than Euclidean. The triple loss function detailed in Section IV shapes the learned constellation such that angular separation between channel symbols corresponds to semantic dissimilarity:
\begin{equation}
    \theta_{ij} = \arccos\left(\frac{\mathbf{c}_i \cdot \mathbf{c}_j}{||\mathbf{c}_i|| \cdot ||\mathbf{c}_j||}\right) \propto \text{Dist}_{\text{sem}}(s_i, s_j),
\end{equation}
where $\text{Dist}_{\text{sem}}(\cdot, \cdot)$ denotes semantic distance. This geometric structure emerges from the combined effect of the three loss components: cross-entropy enforces semantic decision boundaries, MSE constrains the global scale, and cosine similarity aligns angular structure with semantic relationships.

When a transmitted symbol $\mathbf{x} = \mathbf{c}_i$ is corrupted by additive noise $\mathbf{n}$, the received signal becomes $\mathbf{y} = \mathbf{c}_i + \mathbf{n}$. Following the demapper and token recovery process defined in Eq.~\eqref{eq:cosine}, the system performs nearest-neighbor search via cosine similarity. Geometrically, this finds the nearest neighbor on the semantic hypersphere in terms of angular distance.

Under additive noise at moderate SNR, the corrupted vector $\mathbf{y}$ experiences primarily angular perturbation rather than radial displacement, since the decoder's context-aware processing and the final cosine similarity computation discount magnitude variations. Consequently, $\mathbf{y}$ most likely points toward an angular neighbor of $\mathbf{c}$. Due to the learned topological shaping, these angular neighbors are semantic neighbors. Therefore, STC errors are not random but rather bounded angular shifts within local semantic clusters, an contrast to fixed constellations where errors are uniformly distributed across the entire vocabulary.

\subsection{Symbolic Modalities: Semantic Drift}
In symbolic modalities, represented here by text, channel noise induces a phenomenon termed Semantic Drift. Because the STC constellation encodes semantic relationships into angular proximity, decoding errors tend to shift the predicted token to a semantically or syntactically related symbol, often preserving overall sentence coherence despite precision loss.

To quantify the relationship between learned constellation geometry and decoding errors, channel symbols $\mathbf{c}_i, \mathbf{c}_j \in \mathbb{C}^L$ (with $L=32$ in this analysis) were extracted for the target word ``chickens'' and each confusion word using the template sentence: \textit{``Don't count your \{\} before they hatch.''} Each constellation is represented as $\mathbf{r} = [I_1, I_2, \ldots, I_L, Q_1, Q_2, \ldots, Q_L] \in \mathbb{R}^{2L}$, where $I_k$ and $Q_k$ are the in-phase and quadrature components of the $k$-th complex symbol. Two complementary similarity metrics were computed:

\begin{enumerate}
    \item \textbf{Cosine Similarity}: The decoder's native metric, measuring angular distance on the hypersphere:
    \begin{equation}
        S_{\text{cos}}(\mathbf{r}_i, \mathbf{r}_j) = 
        \frac{\mathbf{r}_i \cdot \mathbf{r}_j}
        {\|\mathbf{r}_i\|_2 \cdot \|\mathbf{r}_j\|_2}.
        \label{eq:cosine_sim}
    \end{equation}
    
    \item \textbf{Chamfer Similarity}: Treats each constellation as a 2-D point cloud $\mathcal{P}_i = \{(I_k, Q_k)\}_{k=1}^L$ and computes bidirectional nearest-neighbor distances:
    \begin{equation}
        S_{\text{chamfer}}(\mathbf{c}_i, \mathbf{c}_j) = 
        \frac{1}{1 + d_{\text{chamfer}}(\mathcal{P}_i, \mathcal{P}_j)},
        \label{eq:chamfer_sim}
    \end{equation}
    where $d_{\text{chamfer}}$ is defined as:
        \begin{multline}
        d_{\text{chamfer}}(\mathcal{P}_i, \mathcal{P}_j) = 
        \frac{1}{L}\sum_{\mathbf{p} \in \mathcal{P}_i} 
        \min_{\mathbf{q} \in \mathcal{P}_j} \|\mathbf{p} - \mathbf{q}\|_2 \\
        + \frac{1}{L}\sum_{\mathbf{q} \in \mathcal{P}_j} 
        \min_{\mathbf{p} \in \mathcal{P}_i} \|\mathbf{q} - \mathbf{p}\|_2,
    \end{multline}
\end{enumerate}

To visualize the high-dimensional learned channel symbols, they were projected onto a 2-D manifold using Uniform Manifold Approximation and Projection (UMAP)~\cite{McInnes2018UMAPUM}, a nonlinear dimensionality reduction technique that preserves local neighborhood structure while revealing global cluster topology in Fig.~\ref{text drift}. Under 0~dB SNR, the target word clusters tightly with semantic neighbors. Errors predominantly drift to close synonyms (e.g., ``animals,'' ``birds''), while severe corruption extends to weaker but still coherent associations (e.g., ``children'', ``kids,'' and ``parents''). This clustering confirms that the learned constellation maps semantic relationships into topological proximity.

Fig.~\ref{text metric} quantitatively validates this mechanism. Confusion frequency correlates strongly with cosine similarity in the constellation space. Although stochastic noise causes minor deviations (e.g., ``animals'' ranking), the nearest neighbors are consistently semantic synonyms. Conversely, Chamfer similarity remains uniform, confirming that STC encodes meaning through angular alignment rather than simple geometric shape or magnitude.

\begin{figure*}
  \centerline{\includegraphics[width=0.95\textwidth]{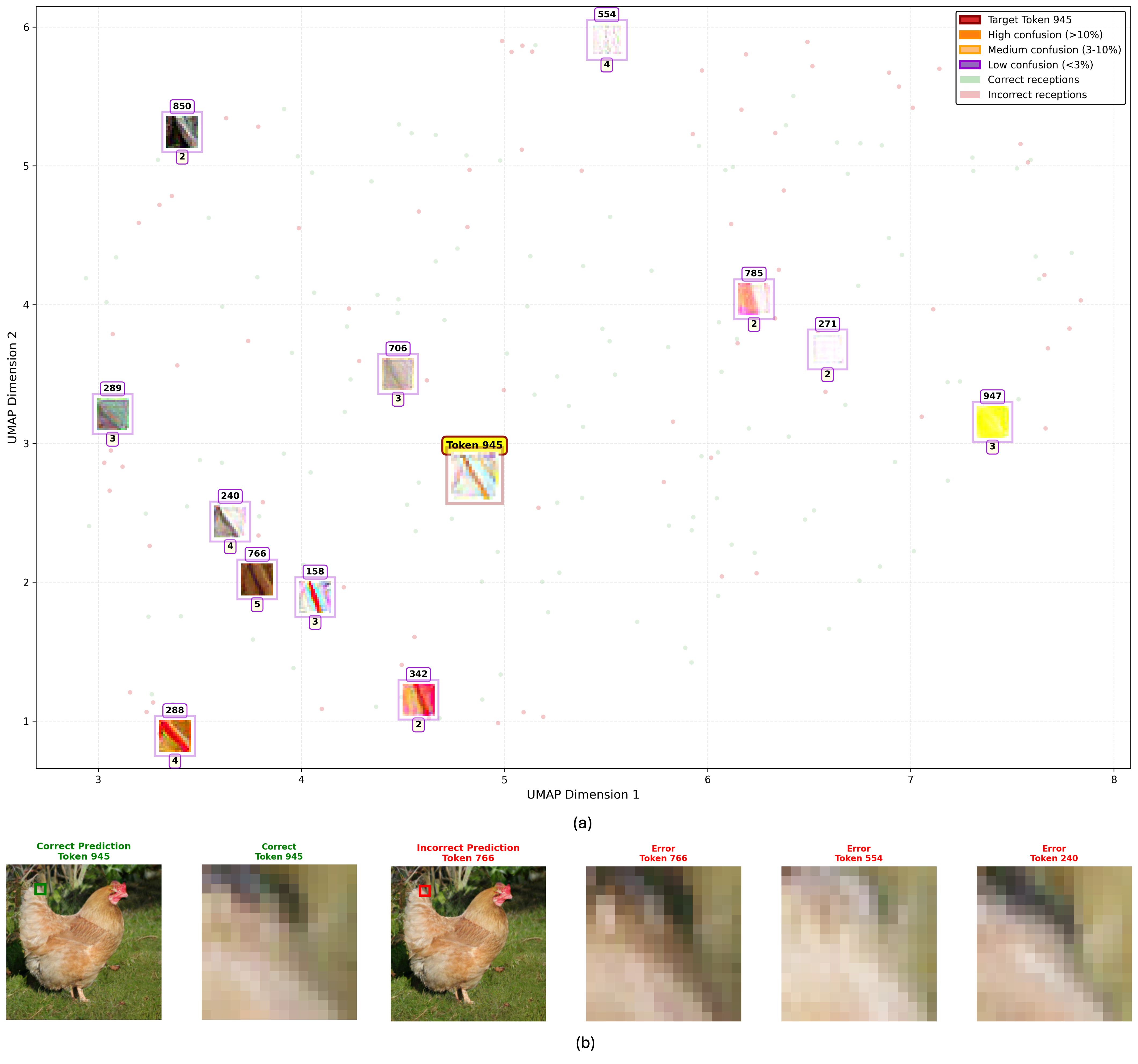}}
  \caption{Visual analysis of structural distortion in image transmission under noisy channel conditions (SNR = 0~dB, $L=32$). (a) UMAP projection of learned channel symbols for target token 945 and its confusion tokens, with each point displaying the corresponding VQ-VAE reconstructed patch thumbnail. Points are colored by confusion rate categories, and labels show confusion frequency. (b) Full image reconstruction comparison showing: correct prediction maintaining target token 945, the isolated correct token patch, incorrect prediction where token 945 was replaced by token 766, 554 and 240 demonstrating graceful degradation. Unlike random bit-flip errors in digital systems with fixed constellations, confused tokens exhibit strong visual similarity to the target, preserving texture and color coherence.}
  \label{image drift}
\end{figure*}

Tables~\ref{tab:context_comparison} and \ref{tab:constellation_similarity_breakdown} further demonstrate context-awareness using the homonym ``bank.'' Under 0~dB SNR, error patterns diverge significantly based on context: financial contexts trigger drifts toward institutional entities, while geographic contexts shift toward landscape terms. High correction rates confirm that STC prioritizes precision, ensuring that inevitable errors manifest only as contextually appropriate synonyms rather than random failures.

\begin{table}[t]
\centering
\caption{Context-dependent semantic drift}
\label{tab:context_comparison}
\begin{tabular}{llcc}
\toprule
\textbf{Context} & \textbf{Confusion Word} & \textbf{Similarity} & \textbf{Corr.} \\
\midrule
\multirow{3}{*}{\parbox{2.5cm}{Financial\\}} 
    & \textit{bank (correct)} & --- & 99.0\% \\
    & court & 0.45 & 0.5\% \\
    & government & 0.37 & 0.5\% \\
\midrule
\multirow{3}{*}{\parbox{2.5cm}{Geographic\\}} 
    & \textit{bank (correct)} & --- & 99.0\% \\
    & river & 0.39 & 0.5\% \\
    & house & 0.35 & 0.5\% \\
\bottomrule
\end{tabular}
\end{table}

\begin{table}[t]
\centering
\caption{Detailed constellation similarity breakdown across contexts.}
\label{tab:constellation_similarity_breakdown}
{%
\begin{tabular}{llcc}
\toprule
\textbf{Context} & \textbf{Pair} & $S_{\text{cos}}$ & $S_{\text{chamf}}$ \\
\midrule
\multirow{2}{*}{Financial} 
    & bank--court & 0.38 & 0.61 \\
    & bank--government & 0.25 & 0.60 \\
\midrule
\multirow{2}{*}{Geographic} 
    & bank--river & 0.36 & 0.58 \\
    & bank--house & 0.20 & 0.54 \\
\bottomrule
\end{tabular}%
}
\end{table}

\subsection{Perceptual Modalities: Structural Distortion}
The perceptual modalities, represented here by images, exhibit an analogous error phenomenon under the STC framework, which is termed Structural Distortion. Instead of random pixel noise, channel errors manifest as coherent alterations in visual patterns that preserve overall image semantics. However, unlike discrete word substitutions in symbolic text, perceptual tokens represent local visual patterns, making distortions continuous in the visual domain rather than categorical in the semantic domain.

To analyze structural distortion, a test image was encoded into a sequence of visual tokens using VQ-VAE tokenization, resulting in a grid of discrete token indices. This sequence was transmitted through the STC system under 0dB noisy conditions and a target patch was selected for analysis. The image was transmitted 200 times, tracking which tokens the target patch confused with due to channel noise.

Critically, the constellation extraction maintains full image context: when computing channel symbols $\mathbf{c}_i \in \mathbb{C}^L$ for each confusion token, only the target patch token was replaced while preserving all other patches at their original values. This ensures that constellation analysis reflects the context-dependent nature of the learned representation, as identical tokens may produce different channel symbols depending on surrounding visual content, which is similar to the context-awareness observed in text.

Using the similarity metrics defined in Eqs.~\eqref{eq:cosine_sim} and~\eqref{eq:chamfer_sim}, cosine and chamfer similarity were computed between the target token's constellation and each confusion token's constellation.

\begin{figure*}[!t]
    \centering
    \includegraphics[width=1\textwidth]{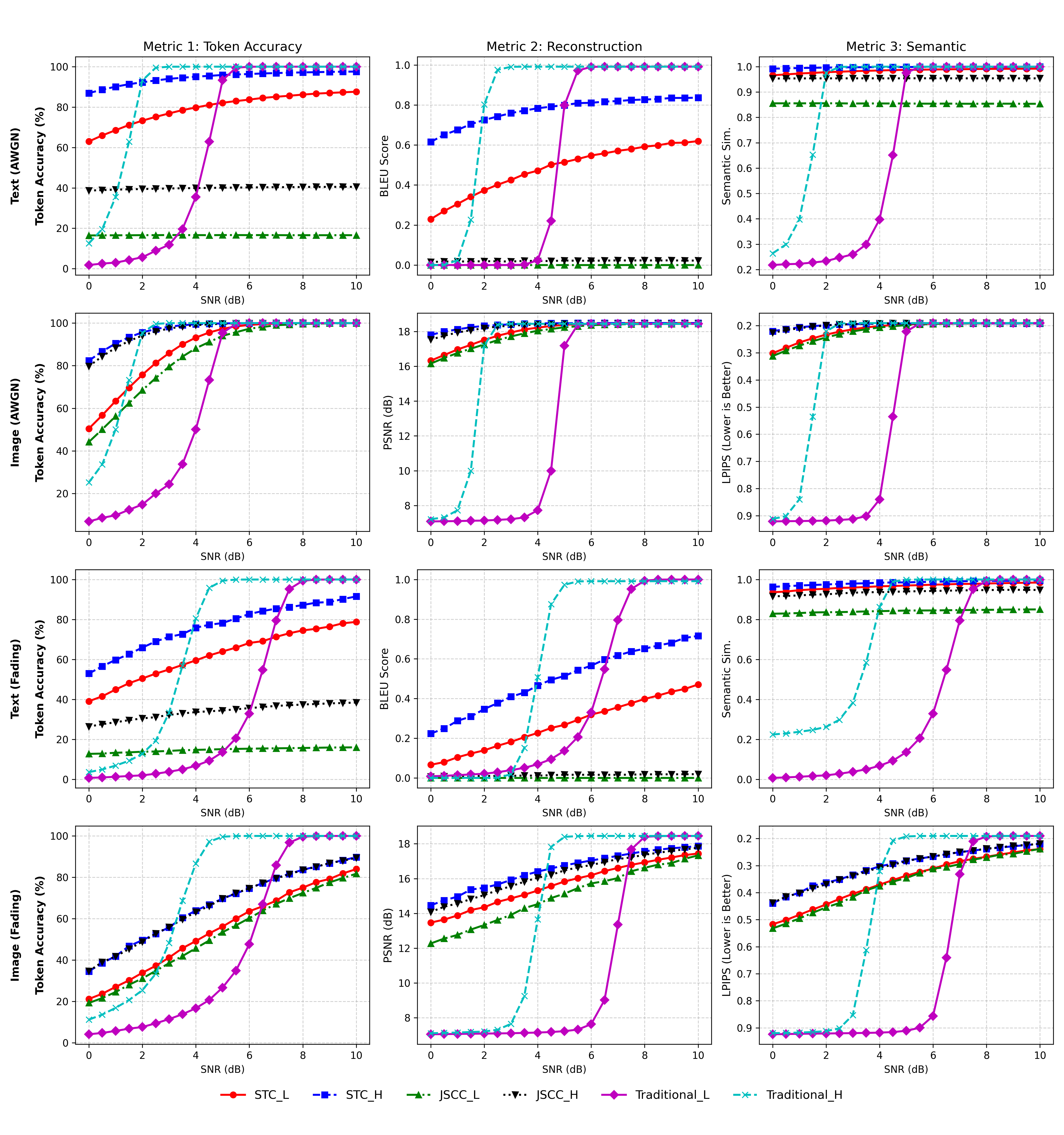}
    \caption{Performance comparison under AWGN and Rayleigh Fading channels for text and image modality. Top two rows: AWGN channel results for text and image. Bottom two rows: Fading channel results. For simplicity, we denote the high-bandwidth configuration with $L=30$ complex symbols per token is denoted as STC\_H and JSCC\_H, 1/2 LDPC + BPSK as Traditional\_H; similarly, the low-bandwidth configuration with $L=15$ complex symbols per token is denoted as STC\_L and JSCC\_L, 1/2 LDPC + QPSK as Traditional\_L. For LPIPS metric, lower values indicate better perceptual similarity, the y-axis is inverted for clarity. STCC demonstrates graceful degradation in low-SNR regimes compared to the ``cliff effect" observed in traditional baselines. }
    \label{fig:snr_results}
\end{figure*}

As visualized in Fig.~\ref{image drift}(a), confusion tokens cluster tightly around patches with similar textures and colors. Quantitative analysis, which is similar to text experiment, also corroborates this: confusion frequency correlates strongly with angular similarity, whereas geometric similarity remains uniform. This confirms that STC encodes perceptual structure primarily through angular topology. Consequently, as shown in Fig.~\ref{image drift}(b), errors result in visually similar substitutions that blend naturally with the surrounding content, validating the system's ability to achieve graceful degradation through topological error shaping.

In summary, structural distortion in the image modality mirrors semantic drift in text: both phenomena arise from the learned hyperspherical constellation topology, which enforces that angular proximity reflects semantic or perceptual similarity. This fundamental property enables STC to achieve graceful degradation under noisy channel conditions, which is a critical advantage over traditional communication systems with fixed constellations that exhibit catastrophic failure modes. The performance benefits of this topological error shaping are quantitatively validated in the next experimental results section.

\section{Experiments}
In this section, the proposed STCC framework is comprehensively evaluated across various channel conditions, semantic tasks, and bandwidth configurations. 

Specifically, the experimental setup, including datasets, comparison schemes, and evaluation metrics, is detailed in Subsection-$A$. Subsection-$B$ analyzes the reconstruction performance of STCC under varying SNR in both AWGN and Fading channels. To validate the practical utility of the decoded semantics, Subsection-$C$ evaluates the performance on downstream tasks, including sentiment analysis and image classification. Subsection-$D$ investigates the trade-off between channel bandwidth usage and system performance. Finally, Subsection-$E$ presents a computational complexity analysis to demonstrate the efficiency of the proposed module.

\subsection{Experimental setup}
\subsubsection{Implementation Details}
To validate the generalizability of the proposed framework, a unified STC architecture and hyperparameter set is maintained for both text and image modalities. Specifically, the STC encoder is configured with $N_b=5$ residual blocks, each with a hidden dimension of $d_h=512$. The decoder consists of 4 Transformer encoder layers with a model dimension of $d_m=512$ and 8 attention heads. All models are trained with a batch size of 32 using the AdamW optimizer \cite{Loshchilov2017DecoupledWD} with a learning rate of $5 \times 10^{-5}$. The $\beta$ values, $\epsilon$, and weight decay for AdamW are set to $(0.9, 0.999)$, $10^{-8}$, and $0.01$, respectively. The weighting hyperparameters in the triple loss function are set to $\lambda_1=10.0$ and $\lambda_2=2.0$. During training, the SNR is randomly sampled from a uniform distribution $\text{SNR}_\text{dB} \sim \mathcal{U}(5, 15)$ to simulate varying channel conditions.

For the text modality, the pre-trained BERT-base model is employed as the foundation model, utilizing its WordPiece tokenizer to generate discrete semantic tokens. It has the codebook size of 30522 $\times$ 768. The system is trained on the Wikitext-103 \cite{Merity2016PointerSM} dataset with a maximum sequence length of 64. For the image modality, the MaskGIT (VQ-GAN) architecture is adopted as the discrete tokenizer and reconstructor. It has the codebook size of 1024 $\times$ 256, which is much smaller than the text modality to validate the generalizability of the STC across different codebook size. The ImageNet \cite{5206848} dataset is utilized, resizing images to $256 \times 256$ pixels before tokenization into a discrete grid of visual indices. In our experiments, we maintain the STCC framework architecture fixed and replace only the internal token processing module STC with other communication schemes to ensure a fair comparison.

\subsubsection{Comparison Schemes}
To demonstrate the superiority of the proposed framework, STC is compared against two traditional separated communication baselines and a Deep JSCC baseline:
\begin{itemize}
    \item \textbf{1/2 LDPC + BPSK}: A standard digital system employing a rate of 1/2 LDPC code for error correction and Binary Phase Shift Keying (BPSK) modulation.
    \item \textbf{1/2 LDPC + QPSK}: A similar digital system utilizing QPSK to increase spectral efficiency.
    \item \textbf{JSCC}: A standard semantic communication baseline without topological shaping. It treats semantic tokens as continuous vectors, directly maps them to channel symbols, and is trained with an MSE loss to minimize reconstruction error.
\end{itemize}
For a fair comparison, all schemes are evaluated under identical average symbol power constraints and bandwidth conditions.

\subsubsection{Evaluation Metrics}
Performance is evaluated using the following metrics: For the text modality, token accuracy (Acc) is used to measure the percentage of correctly decoded tokens; the 4-gram BLEU score \cite{Papineni2002BleuAM} to assess the quality of generated text against reference sentences; and semantic similarity using Eq.~\eqref{eq:cosine_sim} to quantify semantic fidelity. For the image modality, the same token accuracy metric is used, but Peak Signal-to-Noise Ratio (PSNR) and Learned Perceptual Image Patch Similarity (LPIPS) \cite{Zhang2018TheUE} are also included to evaluate reconstruction precision and perceptual similarity, respectively. For the downstream tasks, accuracy is measured on the SST-2 sentiment analysis dataset \cite{Socher2013RecursiveDM} for text and Top-1 accuracy on ImageNet classification for images.

\subsection{Performance under varying SNR}
The proposed STC algorithm is evaluated against the baseline schemes over AWGN and Rayleigh Fading channels. The performance comparisons for text and image modalities are illustrated in Fig.~\ref{fig:snr_results}. For simplicity, the high-bandwidth configuration with $L=30$ complex symbols per token is denoted as STC\_H and JSCC\_H, 1/2 LDPC + BPSK as Traditional\_H; similarly, the low-bandwidth configuration with $L=15$ complex symbols per token is denoted as STC\_L and JSCC\_L, 1/2 LDPC + QPSK as Traditional\_L.

\subsubsection{Robustness in Low-SNR Regimes}
The most significant advantage of STC is observed in the low-SNR regime, where traditional systems suffer from the catastrophic ``cliff effect". As shown in Fig.~\ref{fig:snr_results}, the traditional schemes exhibit a sharp performance drop-off when the channel condition falls below their decoding thresholds. For instance, in the text modality under AWGN, the accuracy of the QPSK baseline collapses from near 100\% to below 10\% as the SNR drops below 4 dB.

In contrast, the proposed STC exhibits graceful degradation. Even at an extremely low SNR of 0 dB, the high-bandwidth STC (STC\_H) maintains a token accuracy of 86.91\% for text and 82.39\% for images, whereas the baselines are effectively non-functional. This confirms that STC successfully converts channel noise into semantic perturbations rather than complete decoding failures. 

Meanwhile, although the perfect reconstruction is unattainable at very low SNR, STC still preserves a significant portion of semantic content, as evidenced by semantic metrics remaining above 0.9 for text and LPIPS scores below 0.3 for images at 0 dB SNR.

\subsubsection{Comparison with JSCC Baseline}
STC also consistently outperforms the standard JSCC baseline across all metrics. In the text modality, the JSCC baseline struggles to map continuous channel outputs back to discrete tokens effectively, plateauing at a significantly lower accuracy due to the huge codebook dimension. The proposed STC, employing topological error shaping, achieves a performance gain of over 40\% in accuracy compared to JSCC. This highlights the necessity of the proposed triple loss objective, which ensures that the received symbols remain clustered around valid semantic tokens in the constellation space.

\subsubsection{Fading Channel Performance}
The resilience of STC is further validated under Rayleigh Fading conditions in Fig.~\ref{fig:snr_results}, bottom row. While fading induces severe fluctuations that disrupt the baselines, STC maintains a steady recovery curve. Specifically, at 0 dB SNR in the fading channel, STC\_H achieves a text similarity above 0.9, significantly surpassing the BPSK baseline, which can hardly work and only achieve about 0.22 similarity.

\subsection{Semantic Task Adaptivity}
\begin{figure}
    \centering
    \includegraphics[width=1\linewidth]{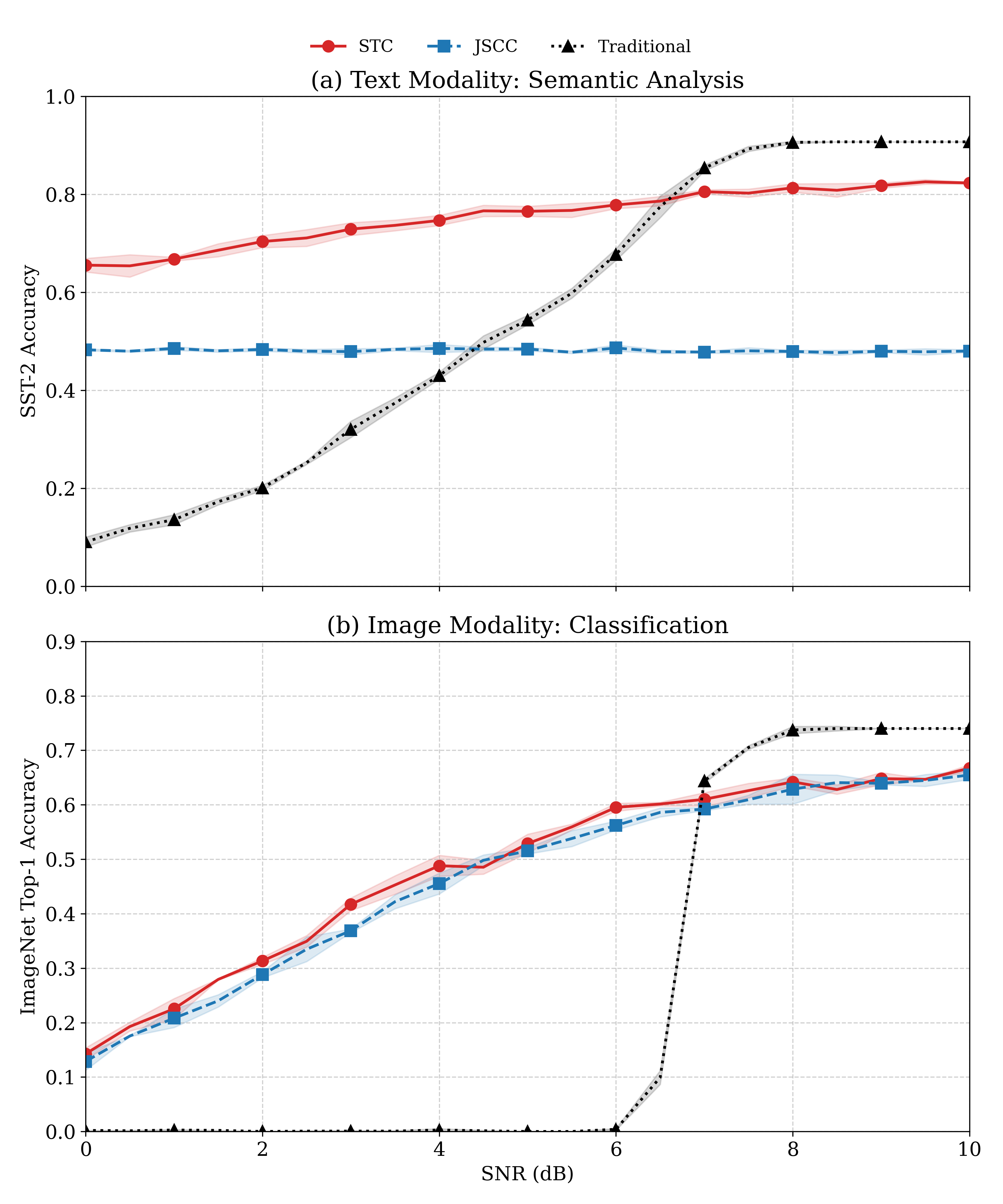}
    \caption{Downstream task performance vs. SNR. (a) SST-2 Sentiment Analysis accuracy for text. (b) ImageNet Top-1 accuracy for images. The shaded regions indicate the standard deviation. STC maintains high task utility in low-SNR regimes where baselines collapse, verifying that decoding errors are topologically bounded and semantically meaningful.}
    \label{fig:task_adaptivity}
\end{figure}

Beyond raw reconstruction metrics, the system's ability to preserve high-level semantic utility for downstream machine intelligence tasks is evaluated. This property is crucial for semantic communication, where the goal is often to convey meaning rather than exact bits. The pre-trained BERT model is employed for Sentiment Analysis (SST-2 dataset) on the decoded text and a ResNet-50 classifier for Image classification (ImageNet) on the reconstructed images. Both experiments are conducted under fading channel conditions. To illustrate the stability of the proposed method, the standard deviation across multiple runs is depicted as a shaded region surrounding the main performance curve. The results are summarized in Fig.~\ref{fig:task_adaptivity}.

\begin{figure*}
    \centering
    \includegraphics[width=1\linewidth]{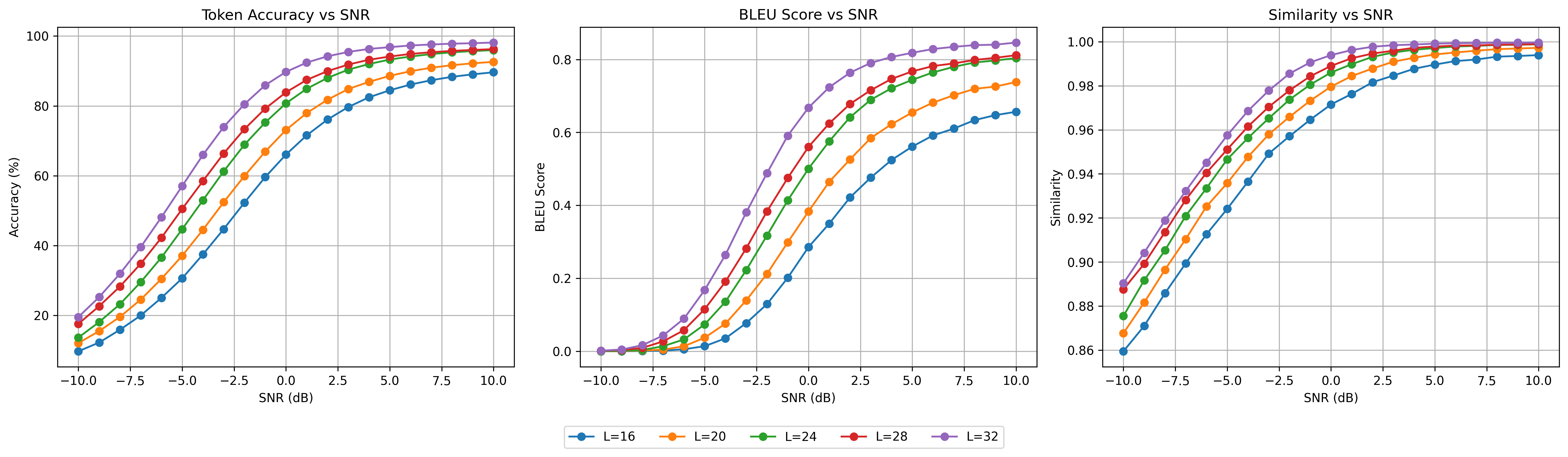}
    \caption{Impact of channel bandwidth cost ($L$) on reconstruction performance. Increasing $L$ significantly enhances robustness, particularly in low-SNR regimes, by allowing for more redundant topological shaping.}
    \label{fig:bandwidth_L}
\end{figure*}

\subsubsection{Text: Sentiment Preservation}
The results in Fig.~\ref{fig:task_adaptivity}(a) highlight a critical advantage of the proposed STC. At a low SNR of 0 dB, STC achieves a sentiment classification accuracy of 65.55\%, significantly outperforming other baselines. This corroborates the theoretical analysis of Semantic Drift: while STC may not recover the exact original words at low SNR, it tends to decode them into semantically related synonyms, thereby preserving the overall sentiment polarity.

In contrast, the traditional baseline yields a negligible accuracy of 9.08\% at 0 dB, as the bit errors corrupt the token indices, resulting in gibberish that is unintelligible to the BERT classifier. Notably, the standard JSCC baseline flatlines near 48\% accuracy across the SNR range. Similar as before, this is because JSCC outputs continuous approximations that, when forcibly discretized, often fail to map to valid or syntactically coherent tokens, confusing the downstream foundation model.

\subsubsection{Image: Semantic Recognition}
A similar trend is observed in the image modality in Fig.~\ref{fig:task_adaptivity}(b). At 0 dB, STC achieves a Top-1 accuracy of 14.27\% on ImageNet, whereas the traditional system fails completely around 0.20\%. As the SNR increases to 5 dB, STC reaches 52.87\%, comparable to the noise-free performance of standard classifiers, while the traditional scheme remains near 0\% until the cliff point at 7 dB. This demonstrates that STC prioritizes the transmission of semantic features such as object shapes and textures over high-frequency pixel details, ensuring that the reconstructed images remain machine-interpretable even under severe channel constraints.

\subsection{Bandwidth-Performance Trade-off ($L$)}
The impact of channel bandwidth usage on system performance is investigated by varying the channel sequence length $L \in \{16, 20, 24, 28, 32\}$. In this context, $L$ represents the number of discrete channel symbols allocated to transmit the semantic tokens, serving as a direct proxy for the bandwidth cost. The performance trade-off is illustrated in Fig.~\ref{fig:bandwidth_L}.

As expected, increasing the bandwidth cost $L$ consistently improves performance across all metrics. However, the critical advantage of higher bandwidth becomes most apparent in the severe noise regime. As shown in Fig.~\ref{fig:bandwidth_L}, under harsh conditions of -5 dB, the model with minimal bandwidth struggles with a token accuracy of only 30.63\%. In contrast, increasing the bandwidth to $L=32$ nearly doubles the performance to 57.05\%. This trend continues at 0 dB, where $L=32$ achieves an accuracy of 89.73\% compared to 66.06\% for $L=16$, representing a substantial gain. This indicates that the proposed STC effectively utilizes the additional bandwidth to expand the geometric spacing between semantic clusters in the channel space, thereby maintaining intelligibility even when the signal power is lower than the noise floor.

Furthermore, the BLEU score demonstrates a similar resilience. At 0 dB, the system with $L=32$ achieves a score of 0.67, significantly outperforming the $L=16$ configuration, which drops to 0.29. This confirms that higher bandwidth allows the decoder to recover finer semantic details and syntactic structures that are otherwise lost in noise. These results demonstrate the flexibility of the STC framework, which can dynamically adjust $L$ to balance transmission reliability and spectral efficiency based on channel capacity.

\subsection{Complexity Analysis}
Finally, the computational complexity of the proposed STC framework is analyzed. The evaluation is conducted on the image modality, utilizing discrete tokens derived from $256 \times 256$ input images via the MaskGIT tokenizer. With the default configuration of $N_b=5$ residual blocks and a hidden dimension of $d_h=512$, the STC module requires only 0.73G Multiply-Accumulate operations (MACs) and contains 2.85M parameters for the combined encoder and decoder.

To put this in perspective, it is compared against a standard ResNet-18 backbone, which typically requires 0.89G MACs and 11.7M parameters. The proposed STC module consumes significantly fewer resources, approximately $18\%$ less computation and $75\%$ fewer parameters than the ResNet-18 baseline. This demonstrates that the STC framework is lightweight and can be integrated into existing semantic token communication systems without introducing significant computational overhead or latency.

\section{Conclusion}
In this paper, the Semantic Token Codebook Communication (STCC), a unified framework that bridges the gap between discrete foundation model tokens and continuous physical channels, was proposed. The STCC framework employs the Semantic Token Codec (STC) as its core engine. By treating the semantic token as a universal interface and employing a novel residual MLP-based mapper with topological error shaping, STC effectively converts random channel noise into bounded semantic variations. 

Theoretical and empirical analyses identified two similar error mechanisms, Semantic Drift in text and Structural Distortion in images, demonstrating that STC preserves high-level semantic utility even when exact reconstruction is impossible. Experimental results confirmed that STC significantly outperforms traditional baselines and existing JSCC methods, particularly in low-SNR regimes where it eliminates the catastrophic ``cliff effect". Furthermore, it was verified that STC maintains high accuracy on downstream tasks, such as sentiment analysis and image classification, while incurring low computational complexity suitable for deployment.

\ifCLASSOPTIONcaptionsoff

  \newpage
\fi

\bibliographystyle{IEEEtran}

\bibliography{IEEEabrv, reference}

\end{document}